\documentclass[preprint]{elsarticle}
\usepackage{amssymb}
\usepackage{amsthm}

\usepackage{lineno}

\journal{}

\begin{document}

\begin{frontmatter}

\title{Role of Nearly-Degenerate Vibrational Modes in Electron Transport through a Molecular Junction}

\author[asc]{Amir Eskandari-asl\corref{cor1}}
\ead{a{\_}eskandari@gate.sinica.edu.tw; amir.eskandari.asl@gmail.com}
\address[asc]{Institute of Atomic and Molecular Sciences, Academia Sinica, Taipei 10617, Taiwan}

\begin{abstract}
We consider an open molecular junction in which electrons are coupled to multi-mode phonons. Using a master equation approach for equilibrated phonons, we show that nearly-degenerate phonon modes can be mapped into an effective single mode, which can result in a noticeable reduction of computational costs. As an illustration, we apply our theory to the OPV-5 molecule and show that we can obtain the consistent conductance spectra while reducing the computational time by orders of magnitude.  
\end{abstract}

%\keywords{master equation; negative differential conductance; Coulomb blockade}%Use showkeys class option if keyword
                              
\end{frontmatter}

%\tableofcontents
Advances in microfabrication and measurement techniques have driven significant progresses in molecular electronics over the past few decades\cite{strachan2005, cui2001, park2002, xu2003, kubatkin2003, dadosh2005, venka2006, moth2009, xin2019, morteza2018}. 
%Pioneering works based on single molecule electronic devices revealed fruitful fundamental quantum physical phenomena, including Kondo effect\cite{park2002,roch2009,parks2010}, Coulomb and Franck-Condon (FC) blockades \cite{park2002,kubatkin2003,leturcq2009,burzuri2014franck}, quantum interference\cite{ballmann2012,vazquez2012,guedon2012}. These groundbreaking experimental findings motivated many theoretical studies to understand nanoscale electron conduction \cite{zimb2011,elke2017,galp2017phot,thoss2018,galp2006opt,galp2008nuc}.
The coupling of the discrete vibrational modes in molecular systems with the transport of electrons brings fruitful quantum physical phenomena, such as negative differential resistance, dynamical switching, current hysteresis, local heating and Franck-Condon blockade. When electron-phonon coupling is present, the charge transfer transition excites quantized molecular vibrations, resulting in modulated conductance spectra that encode the molecular vibrational spectra.

%The interplay between discrete vibrational modes (phonons) and charge transfer transitions can lead to many other vibrationally-resolved phenomena, such as negative differential resistance, dynamical switching, current hysteresis, local heating and Franck-Condon blockade. When coupled to the transport of electrons, the molecular vibrations may appear as the secondary steps in the conductance spectroscopy (current/voltage measurements) or additional equally spaced conductance lines in the charge stability diagrams.\cite{park2002,leturcq2009,burzuri2014franck} In the sequential tunneling regime, the strong electron-phonon coupling leads to a suppression of low-bias conduction which cannot be lifted by varying gate voltage, which is termed as Franck-Condon blockade. 

Many theoretical efforts have been made to investigate the role of quantum vibrations in electron transport through molecular junctions. The most commonly employed approaches are the non-equilibrium Green's functions method\cite{flens2003,mitra2004,galp2006r,hartle2008,erpen2015} for coherent tunneling regime and the master equation approach \cite{may2002,mitra2004,harb2006,timm2008,espo2009,fu2019} for incoherent sequential electron tunneling (SET) regime. Although the side band of conductance spectra reveals the molecular vibrational spectra, nearly degenerate modes may manifest as the same peak and thus indistinguishable. A strongly coupled vibrational mode exhibited in the experimental results of Ref.~\cite{burzuri2014franck} was found to correspond to two weakly coupled nearly-degenerate modes as given in the first principles calculations. It is also hypothesized in Ref.  \cite{mccaskey2015electron} that two nearly-degenerate modes could act as a single mode which corresponds a stronger peak in the conductance spectroscopy. However, an exact relationship between the nearly-degenerate modes and the corresponding effective mode is lacking.

%One important aspect of transport in molecular junctions is the coupling of electrons to the vibrational modes (phonons) of molecule, or,  electron-phonon (e-ph) interaction. Experimentally, strong enough e-ph interaction will result in the appearance of equally spaced lines parallel to the Coulomb diamond edges in the stability diagrams and suppression of the low bias current which is the FC blockade\cite{park2002,leturcq2009,burzuri2014franck}. There are many theoretical works considering e-ph interaction. The most important approaches for these considerations are non-equilibrium Green's functions\cite{flens2003,mitra2004,galp2006r,hartle2008,erpen2015} and master equation \cite{may2002,mitra2004,harb2006,timm2008,espo2009,fu2019}. In most of the experimental setups where the FC blockade is investigated the usage of master equation approach is justified as the coupling of the leads to the molecular junction is weak compared to the phonon energies and electronic level spacing of the molecule, and can be considered to the lowest order. 

%According to the Ref.~\onlinecite{burzuri2014franck}, the measurements reveal a phonon mode with strong coupling to the electronic transport, while the first principle calculations determine two approximately degenerate modes with noticeably lower couplings than the experimental results. In Ref.  \onlinecite{mccaskey2015electron} it is mentioned that two nearly-degenerate modes act as one strongly coupled mode, as their corresponding peaks are indistinguishable in the experimental setup. 

In this work we show analytically that multiple degenerate vibrational modes could be mapped to a single mode whose effective e-ph coupling is the square root of the summation of the squared couplings of the original degenerate modes (see Eq. (\ref{l2})). 
%We prove this equivalence for the equilibrated phonon case, which is relevant for the experimental situation of Ref. \onlinecite{burzuri2014franck}. 
The advantage of this equivalence relationship is two fold. On one hand, the computational cost for molecular systems with multiple degenerate modes could be reduced significantly. On the other hand, one may disregard some nearly-degenerate modes with weak couplings while their combined effects could be of importance to simulate the experimental results. Such underestimations could be avoided using our work.

%Our results not only yield an understanding of the experimental outcomes, but also can be used to noticeably reduce the computation time in the FC calculations for the molecules with nearly-degenerate phonon modes. Based on our work, it is possible to combine the approximately degenerate modes to a single mode and exponentially reduce the number of terms in the summations that yield the transfer rates. Consequently, the time required for the calculations will be drastically reduced. 

In order to specifically show the applicability of our theory we consider the OPV-5 molecule with two anchoring sulfur atoms. We show that for the temperature which is comparable to the energy difference between the nearly-degenerate phonon modes, we can approximate them as degenerate and applying our theory yields a conductance curve very close to the original one. This illustrates the potential applicability of our theory to simulate transport properties of molecules with several nearly-degenerate phonon modes, which would result in correct conductance peaks while noticeably reducing the computational cost.    

This paper is organized as follow. In Sec.\ref{tm} we present a model for studying the effects of multi-mode phonons on the electron transport of molecular junctions within the master equation approach and considering the equilibrated case, prove the equivalence of multi-mode degenerate phonons to an effective model in which the degenerate modes are combined. In Sec.\ref{rd} we consider the molecule OPV-5 and apply our theory to calculate its conductance spectra and compare the results of effective and original models. Specifically, we show that using the effective model the computation time can be reduced by orders of magnitude. Finally, Sec.\ref{conc} concludes our work.               

\section{Theoretical Model}
\label{tm}

\begin{figure}  %[ht!]
\includegraphics[width=8.5cm]{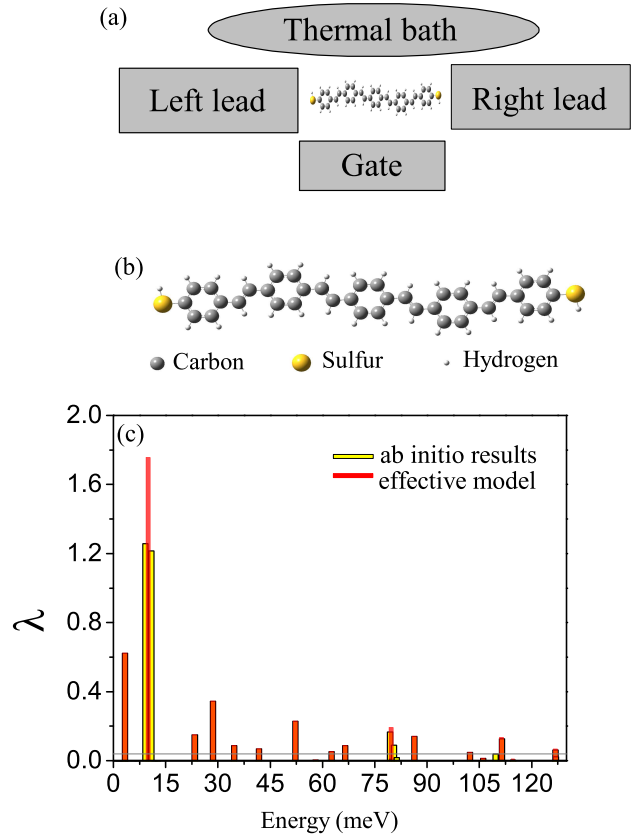}
\caption{\label{sys}(a) Schematic representation of the three-terminal system. A molecular junction is bridged between two metallic leads with which it can exchange electrons, and electrostatically gated by a third lead. A thermal bath is coupled to the molecule so that the phonons can be relaxed. (b) Geometry of the molecule OPV-5 with the anchoring sulfur atoms. (c)Yellow bars: \textit{ab initio} results for the e-ph coupling for the OPV-5 molecule in the $-1 \longleftrightarrow 0$ charge transition. Red bars: the effective e-ph coupling obtained by combination of nearly-degenerate modes (modes with the energy differences at most approximately 2 meV are combined).}
\end{figure}
We consider a three-terminal electronic device in which a single molecule is bridged between two metallic leads (source and drain electrodes) and electrostatically gated by a third lead (gate electrode). A schematic illustration of our model system is shown in Fig \ref{sys}a. The total Hamiltonian $\hat H$ is partitioned as
\begin{eqnarray}
    \hat{H}=\hat{H}_{\mathrm{mol}}+\hat{H}_{\mathrm{env}}+\hat{H}_{\mathrm{coup}}
\end{eqnarray}
where $\hat{H}_{\mathrm{mol}}$ describes the molecule system, $\hat{H}_{\mathrm{env}}$ refers the environment involving electrode electrons and thermal phonons, $\hat{H}_{\mathrm{coup}}$ is responsible for the copuling between the molecular system and the environment.

%can be written as $\hat{H}=\hat{H}_{0}+\hat{H}_{\mathrm{coup}}$, where $\hat{H}_{0}=\hat{H}_{\mathrm{mol}}+\hat{H}_{\mathrm{leads}}+\hat{H}_{\mathrm{th}} $ invovles (our system), while $ \hat{H}_{\mathrm{leads}} $ and $ \hat{H}_{\mathrm{th}} $  are the Hamiltonians of the leads and phonon thermal bath, respectively (the environment). $ \hat{H}_{\mathrm{coup}} $ is for the coupling between the molecule and environment and can be written as $ \hat{H}_{\mathrm{coup}}=\hat{H}_{\mathrm{m-l}}+\hat{H}_{\mathrm{m-th}} $, in which the first (second) term describes the coupling between the molecule and leads (thermal bath). 

The molecular Hamiltonian considers the molecular vibrations and its coupling to electronic transitions. Given in terms of the many-electron basis $\{\vert N,a\rangle\}$, $\hat{H}_{\mathrm{mol}}$ reads 
\begin{eqnarray}
&\hat{H}_{\mathrm{mol}}=\sum_{N,a} \sum_{r=1}^{R} \vert Na\rangle \langle Na \vert  
\biggr(\xi_{Na}+\hbar \omega_{r}\left[ \left( \hat{b}_{r}^{\dagger}-\lambda_{r,Na}\right)\left( \hat{b}_{r}-\lambda_{r,Na}\right)+1/2 \right]  \biggr)    \,.
\label{h-vib}
\end{eqnarray}
$ \vert Na\rangle $ is a many-electron state of the molecule with $N$ labelling the number of excess electrons with respect to the neutral state of the molecule and $a$ referring to the spin multiplicity and energy level. The electronic energy of a many-electron state $ \vert Na\rangle $ is $ \xi_{Na}=E_{Na}-N e V_{G} $ with $E_{Na}$ being the eigenenergy of the electronic Hamiltonian and $ V_{G} $ the applied gate voltage. $ \hat{b}_{r}^{\dagger} $ ($ \hat{b}_{r} $) denotes the creation (annihilation) operator of the $r$th phonon mode out of $R$ modes, whereas $\lambda_{r,Na}$ introduces the displacement along the $r$th mode associated with the electronic state $\vert N,a\rangle$.

The environment consists of the electronic leads and phonon thermal bath, with the respective Hamiltonians $ \hat{H}_{\mathrm{leads}}=\sum_{\alpha=L,R} \varepsilon_{\alpha k \sigma} \hat{c}_{\alpha k \sigma}^{\dagger} \hat{c}_{\alpha k \sigma} $ and $\hat{H}_{\mathrm{th}}=\sum_{q}  \hbar \omega_{q} \hat{b}_{q}^{\dagger} \hat{b}_{q}$, where $ \hat{c}_{\alpha k \sigma}^{\dagger} $ ($ \hat{b}_{q}^{\dagger} $) creates an electron (phonon) with energy $ \varepsilon_{\alpha k \sigma} $ ($ \hbar \omega_{q} $) in the lead $ \alpha $ (thermal bath). The coupling Hamiltonian between the molecule and leads is $ \hat{H}_{\mathrm{m-l}}=\sum_{N,a,b ,\alpha k \sigma} V_{\alpha k \sigma ,NaN-1b} c^{\dagger}_{\alpha k \sigma} \vert N-1b\rangle \langle Na \vert +h.c.$, while the phonons on the molecule are coupled to the thermal bath via the Hamiltonian $\hat{H}_{\mathrm{m-th}}=\sum_{q} \sum_{r} t_{q,r} (\hat{b}_{q}^{\dagger}+\hat{b}_{q}) (\hat{b}_{r}^{\dagger}+\hat{b}_{r})$.

\paragraph*{Polaron transformation.} The state of having $ \nu_{r} $ phonons in the mode $ r $ for all $ r \in \left \lbrace 1,..,R \right \rbrace  $ is shown by $ \vert \mathbf{\nu} \rangle $, and $\vert Na,\mathbf{\nu}\rangle$ means $\vert Na\rangle \otimes \vert\mathbf{\nu}\rangle  $, where $\mathbf{\nu}$ is a $R$ dimensional vector with $ \nu_{r} $s as its elements. Using the operator $ \hat{U}=\mathrm{exp} \left[-\sum_{N,a,r} \lambda_{r,Na} (\hat{b}_{r}^{\dagger}-\hat{b}_{r}) \vert Na\rangle \langle Na \vert\right] $ we exploit polaron transformation to make $ \hat{H}_{\mathrm{mol}} $ diagonal in this basis. The polaron transformation is defined as $ \hat{H}^{\prime}=\hat{U} \hat{H} \hat{U}^{-1}  $ and one can show that $\hat{H}_{\mathrm{mol}}^{\prime}=\sum_{N,a,\mathbf{\nu}} (\xi_{Na}+\epsilon_{\mathbf{\nu}})\vert Na\mathbf{\nu}\rangle \langle Na \mathbf{\nu} \vert +\mathrm{const.}$, where $ \epsilon_{\mathbf{\nu}}=\sum_{r=1}^{R} \hbar \omega_{r} \nu_{r} $, and the constant term is $ \sum_{r=1}^{R} \hbar \omega_{r}/2 $ .

\paragraph*{Master equation.} By performing the polaron transformation, considering the environment to be in the thermal equilibrium, and using the Born-Markov approximations (a straightforward generalization of the single-mode derivation given in Ref. \cite{fu2019}), one can show that the the rate of change of  $ P_{Na,\mathbf{\nu}} $ which is the probability of being in the state $ \vert Na,\mathbf{\nu}\rangle $, can be written as $ \frac{dP_{Na,\mathbf{\nu}}}{dt}=\left.\frac{dP_{Na,\mathbf{\nu}}}{dt}\right\vert_{\mathrm{m-l}} + \left.\frac{dP_{Na,\mathbf{\nu}}}{dt}\right\vert_{\mathrm{m-th}} $, in which the first (second) term on the right side is the contribution from coupling to the leads (thermal bath). The explicit form of the first term is     
\begin{eqnarray}
&&\left.\frac{dP_{Na,\mathbf{\nu}}}{dt}\right\vert_{\mathrm{m-l}}=\sum_{\alpha=L,R}\sum_{N^{\prime},a^{\prime}, \mathbf{\nu^{\prime}}} 
\left( k^{\alpha}_{Na\mathbf{\nu}, N^{\prime}a^{\prime} \mathbf{\nu^{\prime}} } P_{N^{\prime}a^{\prime}, \mathbf{\nu^{\prime}}} - k^{\alpha}_{ N^{\prime}a^{\prime} \mathbf{\nu^{\prime}}, N a\mathbf{\nu} } P_{Na,\mathbf{\nu}} \right) , \qquad 
\label{dpav}
\end{eqnarray}  
where
\begin{eqnarray}
&& k^{\alpha}_{Na\mathbf{\nu}, N^{\prime}a^{\prime} \mathbf{\nu^{\prime}} }=  \Lambda^{\alpha}_{Na\mathbf{\nu}, N^{\prime}a^{\prime} \mathbf{\nu^{\prime}} } \prod_{r=1}^{R} \vert M_{\nu^{\prime}_{r} \nu_{r}} (\lambda_{r}) \vert^{2},\qquad
\label{kav}
\end{eqnarray}
in which
\begin{eqnarray}
&&\Lambda^{\alpha}_{Na\mathbf{\nu}, N^{\prime}a^{\prime} \mathbf{\nu^{\prime}} }=\Gamma^{\alpha} \upsilon_{Na, N^{\prime}a^{\prime}} 
\biggr[  \delta_{N,N^{\prime}+1} f\left(\varepsilon_{Na\mathbf{\nu}, N^{\prime}a^{\prime} \mathbf{\nu^{\prime}}},\mu_{\alpha} \right)+ 
\delta_{N,N^{\prime}-1} \left(1-f\left(\varepsilon_{N^{\prime}a^{\prime} \mathbf{\nu^{\prime}} , Na\mathbf{\nu}},\mu_{\alpha} \right)  \right) \biggr] \qquad
\label{lambv}
\end{eqnarray}
where $ \Gamma^{\alpha} $ is the electron transfer rate between  molecular junction and the lead $ \alpha $, and $  \upsilon_{Na, N^{\prime}a^{\prime}}=\sum_{i} \vert \langle Na \vert \hat{c}_i \vert N^{\prime}a^{\prime} \rangle \vert^2+ \sum_{i} \vert \langle Na \vert \hat{c}^{\dagger}_{i} \vert N^{\prime}a^{\prime} \rangle \vert^2 $, where $ i $ runs over any complete single-particle basis (including spin) for electrons on the molecule. Moreover, $ \delta_{m,n} $ is the Kronecker delta and $f(\varepsilon,\mu)=1/(1+e^{(\varepsilon-\mu)\beta })$ is the Fermi distribution function with $ \beta=1/k_B T $ as the reciprocal temperature, $ \varepsilon_{Na\mathbf{\nu}, N^{\prime}a^{\prime} \mathbf{\nu^{\prime}}}= \xi_{Na}+\epsilon_{\mathbf{\nu}}-\xi_{N^{\prime}a^{\prime}}-\epsilon_{\mathbf{\nu^{\prime}}} $ , and $ \mu_{\alpha}=\mu_0 + e V_{\alpha} $ is the chemical potential of the lead $ \alpha $ (the applied bias voltage is $ V=(\mu_L-\mu_R)/e=V_L-V_R $). Finally\citep{eskandari2019interplay},    %, $ \lambda_{r} $ is the dimensionless electron-phonon coupling of the $ r $th phonon mode, and
\begin{eqnarray}
&& \vert M_{\nu^{\prime}_{r} \nu_{r}} (\lambda_{r}) \vert^{2} =\vert\langle \nu_{r} \vert e^{\lambda_{r} (\hat{b}_{r}^{\dagger}-\hat{b}_{r}) } \vert \nu^{\prime}_{r} \rangle \vert^{2}= 
e^{-\lambda_{r}^{2}} \vert\sum_{j=0}^{\nu_{<}} \frac{(-1)^{j}\lambda_{r}^{2j+\nu_{>}-\nu_{<}} \sqrt{\nu_{r}!\nu_{r}^{\prime}!}}{j!(j+\nu_{>}-\nu_{<})!(\nu_{<}-j)!}\vert^{2} , \qquad
\label{mv}
\end{eqnarray}
in which $\nu_{>(<)}$ is the maximum (minimum) of $\nu_{r}$ and $\nu_{r}^{\prime}$, and  $\lambda_{r} \equiv \vert \lambda_{r,Na}-\lambda_{r,N^{\prime}a^{\prime}} \vert$ is the dimensionless e-ph coupling of the $r$th mode in the transition $Na \longleftrightarrow N^{\prime}a^{\prime}$, in which for simplicity we have dropped the subscripts of $Na$ and $N^{\prime}a^{\prime}$. 

On the other hand, the contribution of the thermal bath is given by\cite{fu2019}
\begin{eqnarray}
&&\left. \frac{dP_{Na \mathbf{\nu}}}{dt}\right\vert_{\mathrm{m-th}}=\sum_{\mathbf{\nu^\prime}} 
\left( k^{th}_{Na \mathbf{\nu},Na \mathbf{\nu^\prime}} P_{Na \mathbf{\nu^\prime}}-k^{th}_{Na \mathbf{\nu^\prime},Na \mathbf{\nu}} P_{Na \mathbf{\nu}} \right), 
\label{dpavth}
\end{eqnarray}
where
\begin{eqnarray}
&&  k^{th}_{Na \mathbf{\nu^\prime},Na \mathbf{\nu}}= \gamma^{th} \sum_{r=1}^{R} \biggr[ \delta_{\nu_r+1,\nu^{\prime}_r}(\nu_r+1) n_{th}(\omega_r)+\delta_{\nu_r-1,\nu^{\prime}_r} \nu_r  (n_{th}(\omega_r)+1) \biggr], \qquad
\end{eqnarray}
in which $ \gamma^{th} $ determines the relaxation rate of phonons to the thermal bath and $n_{th}(\omega_r)=1/(e^{\beta \hbar \omega_r}-1)$.

The electrical current from the lead $ \alpha $ to the molecular junction can be obtained as\cite{fu2019}
\begin{eqnarray}
&&I_{\alpha}= e \sum_{N,a,a^{\prime} } \sum_{\mathbf{\nu} \mathbf{\nu^{\prime}} } \left( k^{\alpha}_{N+1a^{\prime} \mathbf{\nu^{\prime}}, Na\mathbf{\nu} }   - k^{\alpha}_{ N-1 a^{\prime} \mathbf{\nu^{\prime}}, Na\mathbf{\nu}  } \right) P_{Na,\mathbf{\nu}}. \qquad
\label{current}
\end{eqnarray}

\paragraph*{Equilibrated phonons.} Considering the phonon relaxation to the thermal bath, two extreme regimes of unequilibrated and equilibrated phonons are possible. In the unequilibrated regime, the relaxation rate of phonons is much less than the electron transfer rate, i.e., $\gamma^{th} \ll \Gamma^{\alpha} $, so that effects of the thermal bath on phonons can be ignored completely. In the other extreme, the phonon relaxation rate is much larger than the electron transfer rate, i.e., $\gamma^{th} \gg \Gamma^{\alpha} $. For our theoretical consideration we assume phonons to be equilibrated, so that in a very short time scale their population on the molecular junction redistributes according to the thermal equilibrium as
\begin{eqnarray}
P_{Na,\mathbf{\nu}}=e^{-\beta \epsilon_{\mathbf{\nu}} } P_{Na,\mathbf{0} }.
\label{paveq}
\end{eqnarray}
Therefore, instead of the probability distributions $P_{Na,\mathbf{\nu}}$, it is sufficient to investigate
\begin{eqnarray}
P_{Na}=\sum_{\mathbf{\nu}} P_{Na,\mathbf{\nu}} .
\label{pa}
\end{eqnarray}
Using Eqs. (\ref{dpav}) and (\ref{kav}), one can obtain the rate of change of $ P_{Na} $ as
\begin{eqnarray}
&&\frac{dP_{Na}}{dt}=\sum_{\alpha} \sum_{N^{\prime},a^{\prime} } \left( k^{\alpha}_{Na,N^{\prime}a^{\prime}  } P_{N^{\prime}a^{\prime} } - k^{\alpha}_{ N^{\prime}a^{\prime},Na} P_{Na} \right) , \qquad
\end{eqnarray}
where
\begin{eqnarray}
&&k^{\alpha}_{  Na,N^{\prime}a^{\prime} } = \frac{1}{Z} \sum_{\mathbf{\nu},\mathbf{\nu^{\prime}}} \Lambda^{\alpha}_{Na\mathbf{\nu}, N^{\prime}a^{\prime} \mathbf{\nu^{\prime}} } \prod_{r=1}^{R} \vert M_{\nu^{\prime}_{r} \nu_{r}} (\lambda_{r}) \vert^{2} e^{-\beta \epsilon_{\mathbf{\nu^{\prime}}}}, \qquad
\label{keqv}
\end{eqnarray}
in which
\begin{eqnarray}
&&Z=\sum_{\mathbf{\nu}} e^{-\beta \epsilon_{\mathbf{\nu}}}=\prod_{r=1}^{R} \frac{1}{1-e^{-\beta \hbar \omega_{r}}}.
\end{eqnarray}
Moreover, the current is given by
\begin{eqnarray}
&&I_{\alpha}= e \sum_{N,a,a^{\prime} } \left( k^{\alpha}_{N+1a^{\prime}, Na }   - k^{\alpha}_{ N-1 a^{\prime} , Na } \right) P_{Na}. \quad
\label{ieq}
\end{eqnarray}
At low temperatures where $ k_{B} T $ is much lower than all phonon energies, just the terms with $ \mathbf{\nu^{\prime}}=\mathbf{ 0 } $ survive in Eq. (\ref{keqv}) and we have $ Z=1 $, therefore the rate is given by
\begin{eqnarray}
&&k^{\alpha}_{  Na,N^{\prime}a^{\prime} } = \sum_{\mathbf{\nu}} \Lambda^{\alpha}_{Na\mathbf{\nu}, N^{\prime}a^{\prime} \mathbf{ 0 } } \prod_{r=1}^{R} \vert M_{0 \nu_{r}} (\lambda_{r}) \vert^{2} . \qquad 
\label{keq0}
\end{eqnarray}   
It is noteworthy that for a real molecule we may have a huge number of phonon modes which would imply a large Fock space, as a result of which computing this rate can take a very long time. In the following, we show that by mapping the nearly-degenerate modes to a single effective mode, one can reduce the size of Fock space and consequently the computational cost for the numerical calculations.

\paragraph*{Degenerate modes.} We can classify the phonon modes according to their energies and investigate the degenerate manifolds. Suppose that we have $ S $ manifolds of degenerate energies which are determined by the distinct phonon frequencies $\lbrace \omega^{(1)},...,\omega^{(S)} \rbrace$, so that each of the frequencies $\omega_r$ belongs to this set . The order of degeneracy in each of these manifolds is shown by $ R^{(i)} \geq 1 $, the summation of which returns the total number of modes: $ \sum_{i=1}^{S} R^{(i)} = R $. In this notation, the phonon number states are determined by $S$ vectors $ \mathbf{\nu^{(1)}},...,\mathbf{\nu^{(S)}}$, in which the elements of $ \mathbf{\nu^{(i)}} $ are shown by $ \nu_{j}^{(i)} $, where $ i=1,..,S $ and $ j=1,...,R^{(i)} $. In a similar manner, the electron-phonon couplings are shown by $\lambda_j^{(i)}$, where $i$ determines the degenerate manifold and the mode within the $i$-th manifold is determined by $ j $. In short, each subscript $r$ is replaced with a superscript $(i)$ and a subscript $j$, the former determining the corresponding degenerate manifold and the latter showing the mode number within that manifold. With this classification it can be shown that (see Appendix \ref{n-proof})
\begin{eqnarray}
\Lambda^{\alpha}_{Na\mathbf{\nu}, N^{\prime}a^{\prime} \mathbf{ 0 } }=\Lambda^{\alpha}_{Na\mathbf{ n }, N^{\prime}a^{\prime} \mathbf{ 0 } },
\label{lambd}
\end{eqnarray}
in which $\mathbf{ n }$ is a $S$ dimensional vector with the elements $ n^{(i)}=\vert \vert \mathbf{\nu^{(i)}}\vert \vert_{1}=\sum_{j=1}^{R^{(i)}} \nu_{j}^{(i)} $, where $\vert \vert ...\vert \vert_{1}$ is the $L_1-norm$. Note that  $ \Lambda^{\alpha}_{Na\mathbf{ n }, N^{\prime}a^{\prime} \mathbf{ 0 } }$ is actually corresponding to an effective system with $ S $ non-degenerate modes. Rearranging the summation in Eq. (\ref{keq0}), the rate can be written as
\begin{eqnarray}
&&k^{\alpha}_{  Na,N^{\prime}a^{\prime} } =\sum_{\mathbf{ n }} \Lambda^{\alpha}_{N a \mathbf{ n }, N^{\prime}a^{\prime} \mathbf{ 0 } } \sum_{\vert \vert \mathbf{\nu^{(1)}}\vert \vert_{1}=n^{(1)}}...\sum_{\vert \vert \mathbf{\nu^{(S)}}\vert \vert_{1}=n^{(S)}} \prod_{i=1}^{S} \prod_{j=1}^{R^{(i)}} \vert M_{0 \nu_{j}^{(i)}} (\lambda_{j}^{(i)}) \vert^{2}. \qquad
\label{keq1} 
\end{eqnarray} 
The second line of the above expression means we sum over all of the phonon states in which the total number of phonons in each degenerate manifold is fixed, while in the first line we consider all of these fixed values. As it is shown in Appendix \ref{n-proof}, we have
\begin{eqnarray}
&&\sum_{\vert \vert \mathbf{\nu^{(1)}}\vert \vert_{1}=n^{(1)}}...\sum_{\vert \vert \mathbf{\nu^{(S)}}\vert \vert_{1}=n^{(S)}} \prod_{i=1}^{S} \prod_{j=1}^{R^{(i)}} \vert M_{0 \nu_{j}^{(i)}} (\lambda_{j}^{(i)}) \vert^{2}= \prod_{i=1}^{S}  \vert M_{0 n^{(i)}} (\lambda^{(i)}) \vert^{2} , \qquad  
\label{mm}
\end{eqnarray}
where
\begin{eqnarray}
\lambda^{(i)} = (\sum^{R^{(i)}}_{j=1} \lambda_{j}^{(i)}{}^{2})^{1/2}.
\label{l2}
\end{eqnarray}
Substituting Eq.(\ref{mm}) in (\ref{keq1}), we arrive at the important relation
\begin{eqnarray}
k^{\alpha}_{  Na,  N^{\prime}a^{\prime} }=\sum_{\mathbf{ n }} \Lambda^{\alpha}_{Na \mathbf{ n }, N^{\prime}a^{\prime} \mathbf{ 0 } } \prod_{i=1}^{S}  \vert M_{0 n^{(i)}} (\lambda^{(i)}) \vert^{2} , \qquad
\label{krel} 
\end{eqnarray} 
which is the same rate as the case of having $ S $ non-degenerate phonon modes, each of which with the frequency $ \omega^{(i)} $ and coupling $ \lambda^{(i)}  $, given by the Eq. (\ref{l2}) (compare with Eq. (\ref{keq0})). As a result, the electronic population distributions and currents are also the same for the two cases. Consequently, we proved that for the equilibrated phonons at low temperatures, as long as the transport properties are considered a molecular junction with several degenerate phonon modes is equivalent to an effective system with non-degenerate modes whose couplings are the square root of the sum of the squared couplings of the degenerate modes of the original molecule (Eq.(\ref{l2})). 

This finding can be physically understood by noting that the square of e-ph coupling is the Huang-Rhys factor, which roughly determines the average number of phonons excited by the transition from one molecular state to another. According to Eq. (\ref{l2}), the Huang-Rhys factor for an effective mode is the summation of the Huang-Rhys factors of the original degenerate modes. Consequently, the total number of excited phonons with a specific energy remains the same in the original and effective models. Our finding suggests that in the electron transport simulations this total number matters.

\paragraph*{Nearly-degenerate modes.} In most of the real molecules we have nearly-degenerate modes instead of the exact degeneracy, i.e., in each nearly-degenerate manifold, we have $R^{(i)}$ phonon modes with the frequencies $\omega_1^{(i)},...,\omega_{R^{(i)}}^{(i)}$. Our derivation is still applicable to such cases if these frequencies are indistinguishable. Noting that the key usage of degeneracy was the factorization that allowed us to go from Eq. (\ref{keq0}) to Eq. (\ref{keq1}), we can say as long as the difference between two phonon energies is at most comparable with the thermal energy, $k_B T$, this factorization is valid and we can approximate the modes as being degenerate. Therefore, one can do the numerical calculations for an effective system in which each group of the nearly-degenerate modes are combined into a single mode with a coupling that is given by the Eq. (\ref{l2}). The energy of this equivalent effective mode can be anywhere in the range determined by the nearly-degenerate modes, however, based on the equivalent coupling strength and the physical meaning of the Huang-Rhys parameters, we suggest a weighted average of the nearly-degenerate energies with the Huang-Rhys parameters as the weight factors. It should be noted that there are other possible origins of the nearly-degenerate modes to be indistinguishable in experimental situations, such as having energy differences less than the level broadening or other experimental uncertainties that are beyond the scope of our model.

One important usage of our results is to reduce the computational cost of transport simulations. In a variety of molecules, we encounter nearly-degenerate phonon modes. Considering all of such modes may require a long time for numerical calculations, as a large number of terms should be summed over to calculate the rates. However, using our effective model with the nearly-degenerate modes combined, the phonon Fock space will be shrunk. Consequently, the number of these terms will be reduced drastically which would result in shortening the time needed to do the numerical simulations. 

Another important issue to be addressed is ignoring the weakly coupled modes. In the cases that the \textit{ab initio} calculations indicate several degenerate modes with relatively weak couplings, one may ignore the importance of the combined effects of the modes and loose some features in their simulations. However, knowing that those degenerate modes are equivalent to a single-mode with an effective coupling given by Eq. (\ref{l2}) can avoid such issues.  

\begin{figure}  %[ht!]
\includegraphics[width=8.cm]{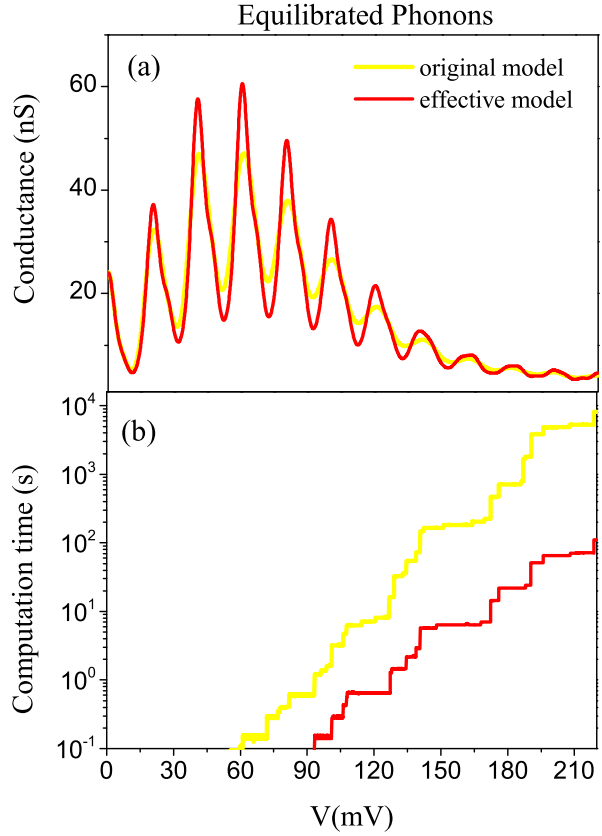}
\caption{\label{gv}  (a) Conductance as a function of the applied bias voltage for the molecule OPV-5 for the equilibrated phonon case, for the original model (yellow) and effective one (red ). For these calculations all of the modes with e-ph coupling less than 0.04 are ignored (bellow the gray line in Fig \ref{sys})c. (b) Conductance computation time (non-cumulative) as a function of the applied bias voltage for the original and effective models. The times smaller than 0.1s are not shown in this figure.  The tunneling rate is $\hbar \Gamma_L=\hbar \Gamma_R=0.05 $ meV, the temperature is $k_B T=1$ meV ,and the value of gate voltage is chosen such that the charged state is aligned with the neutral state at zero bias. }
\end{figure}

\section{Results and discussion} 
\label{rd}
In order to investigate the applicability of the nearly-degenerate approximation for a real molecule, we consider the OPV-5 molecule with two anchoring sulfur atoms, the geometry of which is shown in Fig \ref{sys}(b). Using the Gaussian 09 software at the B3LYP\cite{becke1993} level and with the cc-pVDZ basis set, we did the geometry optimization and obtained the phonon mode energies. Moreover, we performed the Duschinsky transformation\cite{duschinsky} using the DUSHIN\cite{DUSHIN} code to calculate their e-ph couplings for the $-1 \longleftrightarrow 0$ charge transition , as plotted with the yellow bars in Fig \ref{sys}(c). Also shown in the Fig \ref{sys}(c) with the red bars, are the effective couplings obtained by combining nearly-degenerate modes with the energy differences which are at most approximately equal to 2 meV. This approximation is justified as we consider the temperature to be $k_B T=1$ meV. We consider a gate voltage that at zero bias aligns the charged and neutral states, and assume a symmetric applied bias voltage. The tunneling rates are also considered symmetric with the values of $\hbar \Gamma_L=\hbar \Gamma_R=0.05 $ meV. In Fig.\ref{gv}(c) we plot the conductance, $dI/dV$, as a function of the applied bias voltage in the equilibrated phonon case, for the original model (yellow) and effective one (red). In both of these calculations all modes with the e-ph coupling less than 0.04 are ignored (i.e., those with the couplings bellow the gray line in Fig \ref{sys})c).  As it is seen in this figure, the two curves are consistent which shows the validity of our effective model. However, the computation time for the two models are drastically different, as it is shown in Fig \ref{gv}(b), where we plot the conductance computation time consumed by a single CPU as a function of bias voltage for the two models. Specially, for the higher bias voltages, the computation time for the effective model can be few orders of magnitude less than that of the original model. Note that one has to do the calculations for each bias voltage and these times should be integrated to give the consumed time, which means the total time can be reduced from few weeks to few hours by using the effective model instead of the original one.

\section{Conclusions}
\label{conc}
In conclusion, we considered an open molecular junction whose electrons are coupled to multi-mode phonons, using a master equation approach. We theoretically investigated the equilibrated phonon case in which the phonons are strongly coupled to a thermal bath so that they maintain the thermal equilibrium. We proved that when the temperature is low compared to the phonon energies, (nearly-) degenerate phonon modes can be combined to an effective single-mode with the coupling given by the square root of the summation of the squared couplings of the original system (Eq. (\ref{l2})).

 In order to show the applicability of our theory in an illustrative example, we considered the OPV-5 molecule with two anchoring sulfur atoms, and showed that the conductance calculated using the effective combined model is consistent with the results obtained by considering the original modes, while the computation time for the former is drastically less than the latter. This effective model can reduce the computation time for transport simulations, as nearly-degenerate modes are combined so that less time is needed for calculating the transition rates in the reduced effective Fock space. 

One open question for a future work is the case where some phonon peaks are indistinguishable at low temperatures as a result of level broadening. As an example, in Ref. \cite{burzuri2014franck} one conductance peak determining a strong e-ph coupling is reported in the experimental results while the \textit{ab initio} calculations reveal two nearly degenerate modes at the corresponding energy. Even though in their setup the temperature is low, the level broadening is comparable to the energy difference between the modes which makes them behave as a single effective mode.\cite{mccaskey2015electron} 

\section*{ACKNOWLEDGMENT}
We acknowledge useful discussions with Liang-Yan Hsu, Bo Fu, and Qian-Rui Huang.

\appendix

\section{Proof of Eqs. (\ref{lambd}) and (\ref{mm})}
\label{n-proof}
In order to prove Eq. (\ref{lambd}) we note that 
\begin{eqnarray}
&&\epsilon_{\mathbf{\nu}}=\sum_{r=1}^R \hbar \omega_{r} \nu_r =\sum_{i=1}^{S} \hbar \omega^{(i)} \sum_{j=1}^{R^{(i)}} \nu_{j}^{(i)}, \qquad
\label{epva}
\end{eqnarray}
by defining $n^{(i)}$ as the $L_1-norm$ of $\mathbf{\nu^{(i)}}$, the left hand side of Eq. (\ref{epva}) can be written as $\epsilon_{\mathbf{n}} \equiv \sum_{i=1}^{S} \hbar \omega^{(i)} n^{(i)}$, which corresponds to the phonon energy of an effective system with $S$ phonon modes with the occupation numbers of $n^{(i)}$. Substituting this into Eq. (\ref{lambv}) (with $\mathbf{\nu^\prime}=\mathbf{0}$) one arrives at the Eq. (\ref{lambd}). 

For proving Eq. (\ref{mm}) we notice
\begin{eqnarray}
&&\sum_{\vert \vert \mathbf{\nu^{(1)}}\vert \vert_{1}=n^{(1)}}...\sum_{\vert \vert \mathbf{\nu^{(S)}}\vert \vert_{1}=n^{(S)}} \prod_{i=1}^{S} \prod_{j=1}^{R^{(i)}} \vert M_{0 \nu_{j}^{(i)}} (\lambda_{j}^{(i)}) \vert^{2}= \nonumber\\
&&\prod_{i=1}^{S} \left[\sum_{\vert \vert \mathbf{\nu^{(i)}}\vert \vert_{1}=n^{(i)}}  \prod_{j=1}^{R^{(i)}} \vert M_{0 \nu_{j}^{(i)}} (\lambda_{j}^{(i)}) \vert^{2} \right] 
\label{lhs}
\end{eqnarray}

Recalling from Eq. (\ref{mv}) that $ \vert M_{0 \nu} (\lambda) \vert^{2}=\frac{\lambda^{2 \nu}}{\nu!} e^{-\lambda^2 }$, each of the terms in the bracket can be written as
\begin{eqnarray}
&&\sum_{\vert \vert \mathbf{\nu^{(i)}}\vert \vert_{1}=n^{(i)}}  \prod_{j=1}^{R^{(i)}} \vert M_{0 \nu^{(i)}_{j}} (\lambda^{(i)}_{j}) \vert^{2}= \nonumber\\
&&\mathrm{exp}\left(-\sum_j \lambda^{(i) 2}_{j}\right) \sum_{\vert \vert \mathbf{\nu^{(i)}}\vert \vert_{1} = n^{(i)} } \prod_{j=1}^{R^{(i)}} \frac{\lambda_{j}^{(i) 2 \nu^{(i)}_j}}{\nu^{(i)}_j !} .
\label{srw}
\end{eqnarray}
Using Eq. (\ref{l2}) as the definition of $\lambda^{(i)}$, one can check that
\begin{eqnarray}
\left(\lambda^{(i)} \right)^{n^{(i)}}=n^{(i)}! \sum_{\vert \vert \mathbf{\nu^{(i)}}\vert \vert_{1}=n^{(i)}} \prod_{j=1}^{R^{(i)}} \frac{\lambda_{j}^{(i) 2 \nu^{(i)}_j}}{\nu^{(i)}_j !}.
\label{ln}
\end{eqnarray}
Substituting Eqs. (\ref{srw}) and (\ref{ln}) into Eq. (\ref{lhs}) results in Eq. (\ref{mm}).

\bibliographystyle{model1-num-names}
\bibliography{mp-016.bib}

\end{document}